\documentclass[prb, twocolumn, showkeys,showpacs,amsmath,amssymb]{revtex4}
\usepackage{amssymb}
\usepackage[T2A]{fontenc}

\usepackage{epsf,graphicx}
\begin{document}
\title{Calculation of the P-T phase diagram and tendency toward decomposition in equiatomic TiZr alloy}
\author{V.Yu.~Trubitsin, E.B.~Dolgusheva}
\date{}
\email{tvynew@otf.pti.udm.ru}
\affiliation{Physical-Technical Institute,
Ural Branch of Russian Academy of Sciences,
132 Kirov Str.,426001 Izhevsk, Russia}

\date{\today}

\begin{abstract}
Electronic, structural and thermodynamic properties of the equiatomic alloy TiZr are calculated within the electron density functional theory and the Debye-Gr\"{u}neisen model. The calculated values of the lattice parameters $a$ and $c/a$ agree well with the experimental data for the  $\alpha$, $\omega$ and $\beta$ phases. The $\omega$ phase is shown to be stable at atmospheric pressure and low temperatures; it remains energetically preferable up to $T=600$~K. The $\alpha$ phase of the TiZr alloy becomes stable in the range $600$~K$<T<900$~K, and  the $\beta$ phase at  temperatures above $900$~K. The constructed phase diagram qualitatively agrees with the experimental data available. The tendency toward decomposition in the equiatomic alloy $\omega-$TiZr is studied. It is shown that in the ground state the $\omega$ phase of the ordered equiatomic alloy TiZr has a tendency toward ordering, rather than decomposition.
\end{abstract}
\maketitle

\section *{Introduction}

It has been experimentally found that the TiZr system is characterized by full solubility of its constituents. As in pure titanium and zirconium, three phases ($\alpha$, $\beta$ and $\omega$) are observed in the TiZr alloy \cite{bashkin,bashkin2,aksenenkov}. The structural $\alpha \to \beta$ transformations of the equiatomic alloy were extensively studied in Ref. \cite{bashkin} by the differential thermal analysis (DTA) at temperatures up to $1023$~K, and pressures up to $7$~GPa.   It was found that the $\beta \to \alpha$ transition temperature, being equal to $852$~K at atmospheric pressure, decreases with pressure down to the triple equilibrium point of the $\alpha$, $\beta$ and $\omega$ phases ($P_{tr}=4.9 \pm 0.3$~GPa, $T_{tr}=733 \pm 30$~K). At pressures above the triple point the $\beta$ phase transforms immediately to the $\omega$ phase with a light positive slope of the equilibrium line. If a sample is cooled to room temperature at a pressure of $6$~GPa, and then unloaded, one can obtain at atmospheric pressure a metastable $\omega$ phase which on heating transforms into an $\alpha$ phase in the temperature interval from $698$~K to $743$~K. Cooling of the $\beta$ phase in the pressure range $2.8$ - $4.8$~GPa results in the formation of a two-phase mixture of a stable $\alpha$ and a metastable $\omega$ phase. The structural $\alpha \to \omega$ transformations in the TiZr alloy were studied in detail in Ref. \cite{aksenenkov}. Investigating TiZr samples under shear-strain conditions at pressures up to $9$~GPa at temperatures $300$ and $77$~K, the authors arrived at the conclusion that in equiatomic TiZr the equilibrium $\alpha \to \omega$ boundary is situated on the P-T diagram at $6.6$~GPa. In the same paper the phase diagram of TiZr was constructed in the regular-solution approximation, and the triple point parameters were calculated ($P=8.5$~GPa, $T=693$~K). As may be seen, these values differ substantially from those of Ref. \cite{bashkin}. 

Detailed studies performed in Ref. \cite{Dmitriev} have shown that in the region of high pressures and temperatures there exist two $\omega$ phases ($\omega$ and $\omega_1$) that differ in atomic volume by about $14$\%. The authors suggested the existence of a  isostructural phase transition $\omega - \omega_1$ connected with changes in the electron structure of the alloy. They supposed that the large difference in the atomic volume between the two phases points to the existence of an s-d electronic transition in $\omega-$TiZr. Later, phase separation of a hexagonal TiZr $\omega$ phase was experimentally detected in Ref. \cite{Bashkin-2008}. The $\omega \to \omega_1+\omega_2$ decomposition  was revealed after a prolonged heat treatment at $P=5.5 \pm 0.6$~GPa and $T=440 \pm 30^0$~C. It was supposed that in a wide concentration range at pressures above the triple equilibrium point, the $\omega$ phase may exist in the $Ti_xZr_{1-x}$  alloy only as a metastable one that persists due to low diffusive mobility of its constituents. The decomposition of the $\omega-$TiZr solid solution into two $\omega$ phases of different structure was used as an alternative explanation for the experimental results obtained in Ref. \cite{Dmitriev}.

Up to now the electron structure and structural transformations of the equiatomic alloy TiZr have never been calculated. Below we present the results of our theoretical calculations of electronic, structural and thermodynamic properties of the equiatomic alloy TiZr performed within the framework of the electron density functional theory and the Debye-Gr\"{u}neisen model. The tendency of the ordered equiatomic alloy $\omega-$TiZr to decompose is also investigated.

\section {Calculation technique}

The electron structure and total energy were calculated by the scalar relativistic full-potential linearized augmented-plane-wave (FPLAPW) method, using the WIEN2K package \cite{Wien2k}. To ensure the desired accuracy of the total energy calculation, the number of plane waves was defined by the condition $RK_{max}=7$, the total number of $k$-points in the Brillouin zone was equal to 3000, 3000, 600 for the $\beta$, $\alpha$ and $\omega$ phase, respectively. The total and partial densities of states were obtained by a modified tetrahedron method \cite{tetrahedron}. The atomic radii were the same for all phases and pressures: $2.42$~a.u.  for Zr, and  $2.26$~a.u.  for Ti.

\begin{figure}[!tbh]
\begin{center}
 \begin{minipage}{3.0in}
 \epsfxsize=0.9in
 \epsfysize=0.9in
 \epsfbox{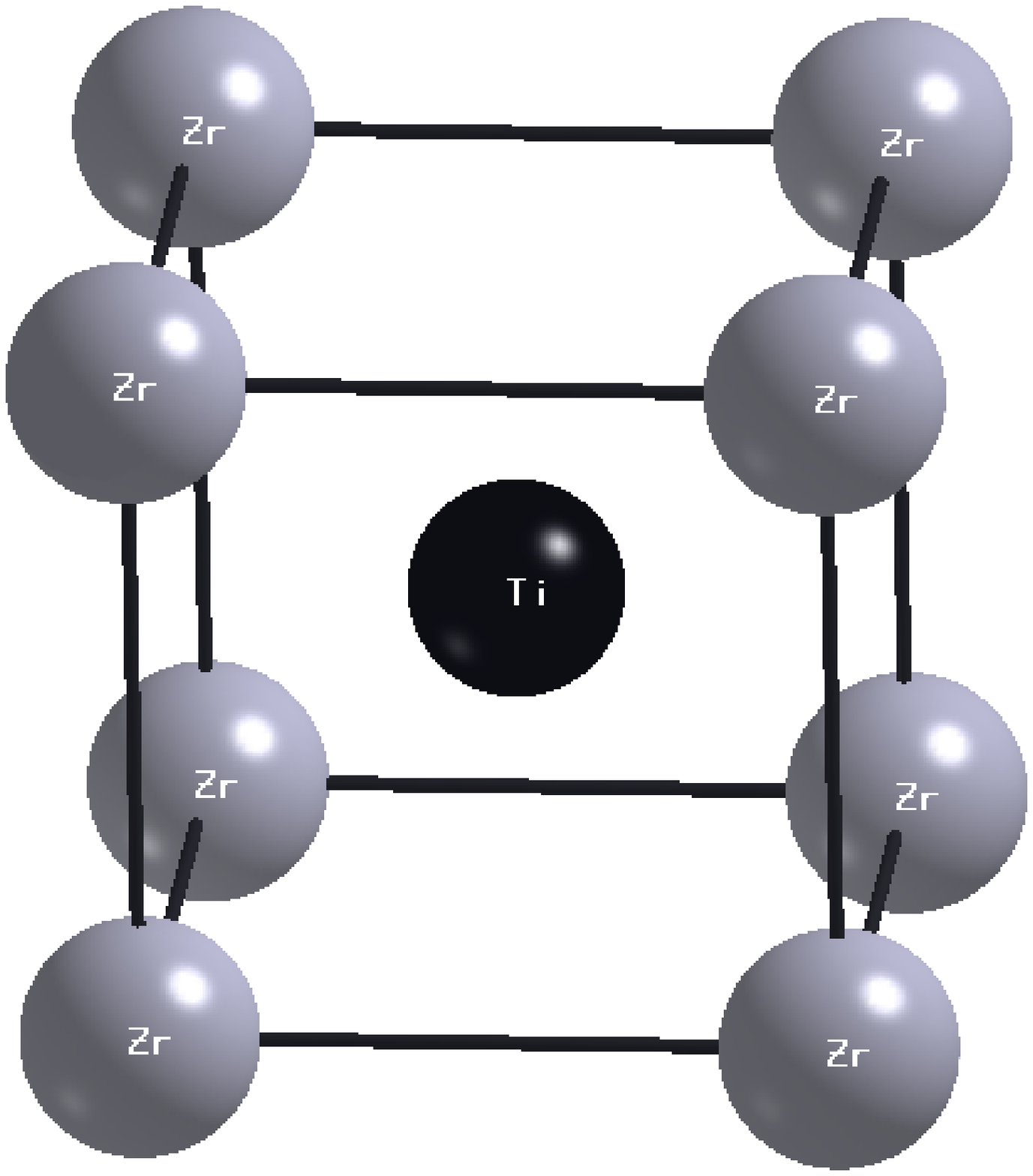}
 \epsfxsize=0.9in
 \epsfysize=0.9in
 \epsfbox{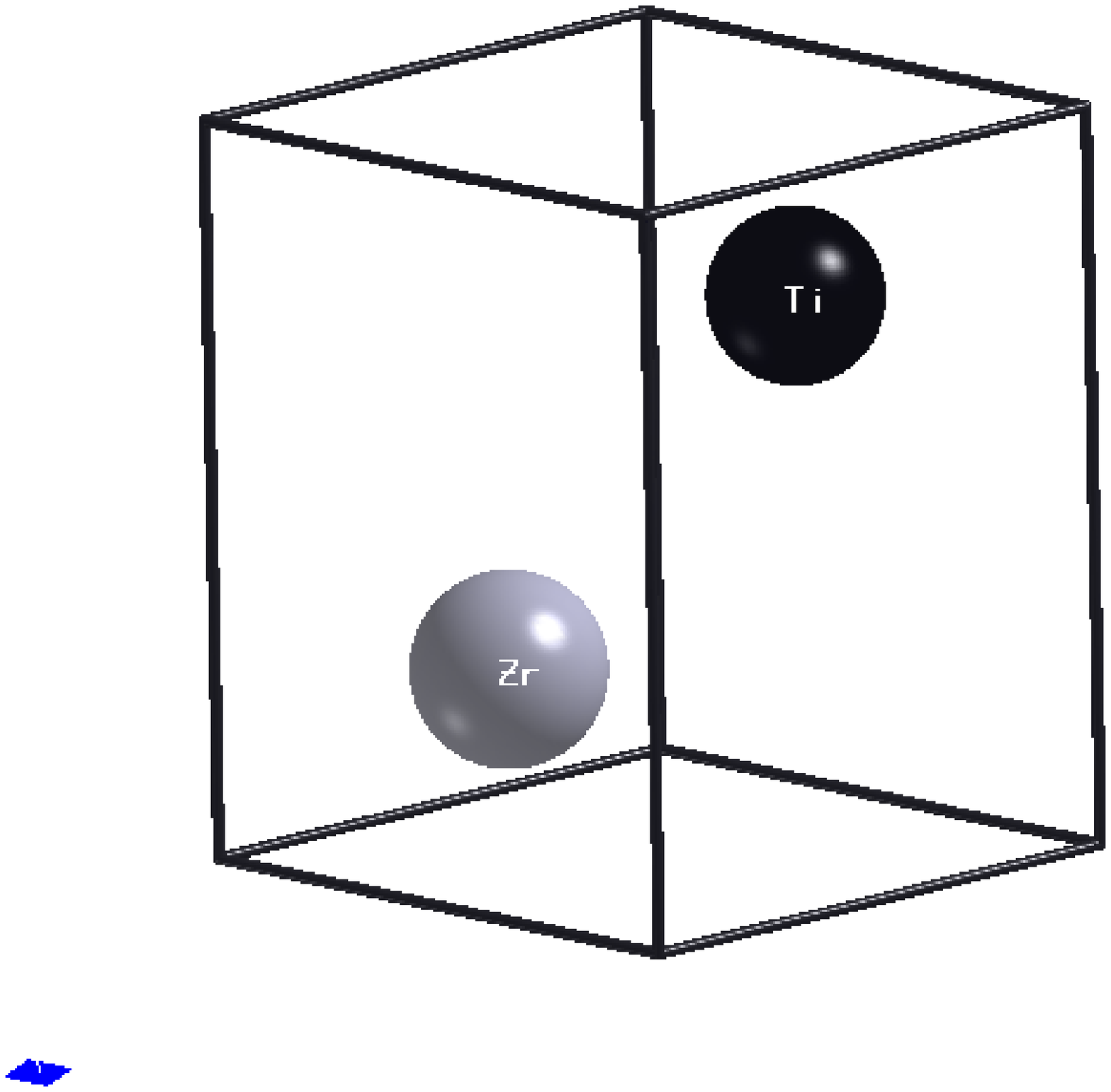}
 \epsfxsize=0.9in
 \epsfysize=0.9in
 \epsfbox{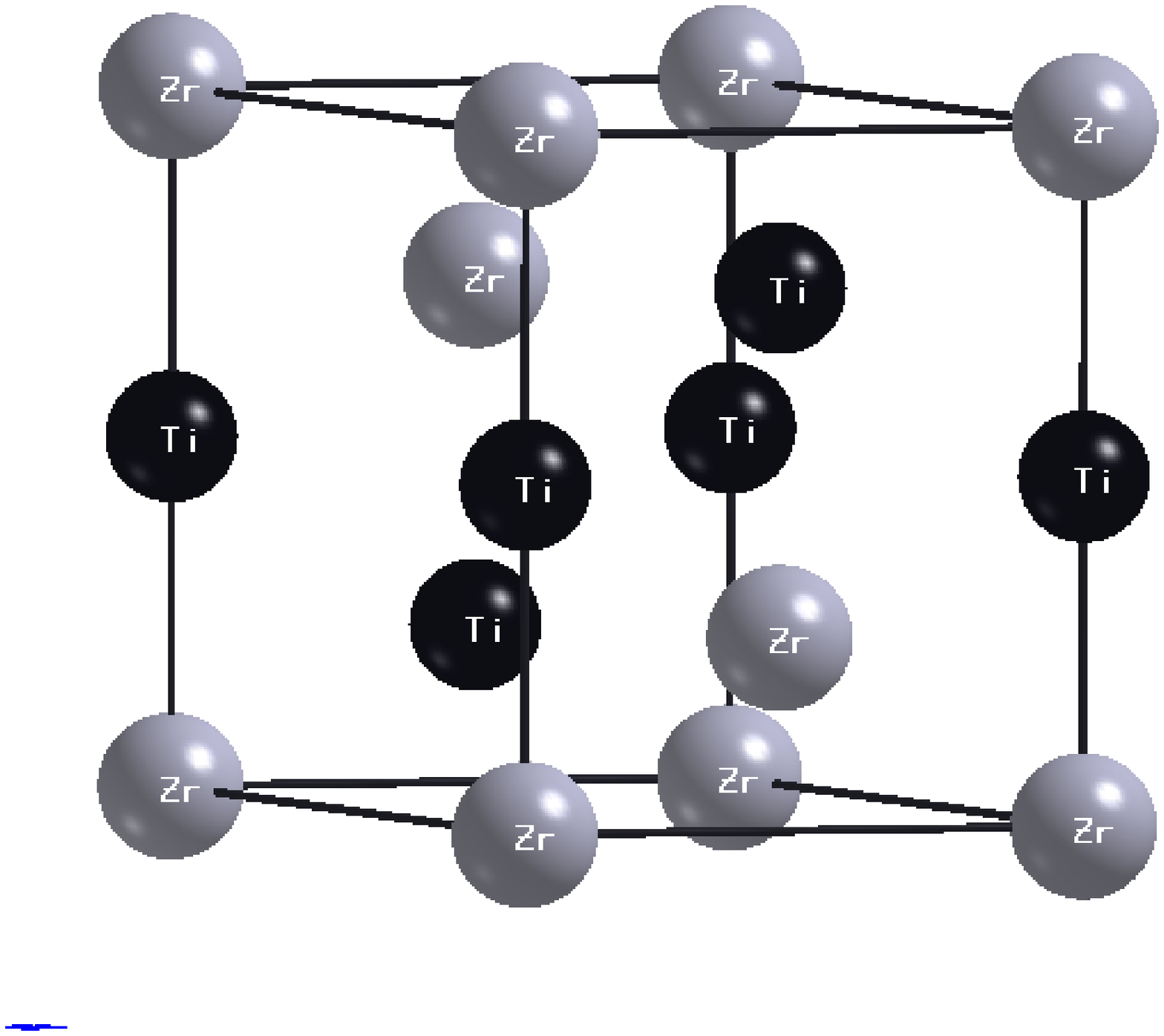}
 \end{minipage}
\caption{Crystal structure of the $\beta$, $\alpha$ and $\omega$ phases of equiatomicTiZr. The Zr atoms are grey, the Ti atoms are dark.}
\label{Fig: TiZr_Cristal_structure}
\end{center}
\end{figure}

In Fig.\ref{Fig: TiZr_Cristal_structure} are shown the crystal structures of the $\beta$, $\alpha$, and $\omega$ phases of the equiatomic alloy TiZr used in the calculation. It is seen that the $\beta$ phase was represented by a structure of the CsCl type with Ti atoms at the cube sites and a Zr atom at the center. The $\alpha$ phase had a hexagonal close-packed lattice in which one atom was Zr, the other Ti. Finally, to describe the $\omega$ phase we used a hexagonal lattice with 6 atoms per cell (an $\omega$ structure doubled along the $z$ axis). The atomic arrangement and species in this case were chosen as follows:
$(0,0,0)$ - Zr,
$(\frac{1}{3},\frac{2}{3},\frac{1}{4})$ - Ti,
$(\frac{2}{3},\frac{1}{3},\frac{1}{4})$ - Zr,
$(0,0,\frac{1}{2})$ - Ti,
$(\frac{1}{3},\frac{2}{3},\frac{3}{4})$ - Zr,
$(\frac{2}{3},\frac{1}{3},\frac{3}{4})$ - Ti.
Thus, the $\beta$ and $\alpha$ phases were represented by layers of Ti and Zr alternating along the $z$ axis, and in the $\omega$ phase  layers of Ti and Zr were separated by a mixed Ti-Zr layer.

In the hexagonal structures the ratio $c/a$ was optimized for the experimental volume values. In the following, when calculating the volume dependence of the total energy, the $c/a$ ratio was considered to be constant. In Table \ref{table:TiZ_lattice_parameter} are listed the calculated and experimental equilibrium values of the TiZr lattice parameters.

\begin{table}[!tbh]
\begin{center}
\caption{Equilibrium values of the TiZr lattice parameters in atomic units}
\begin{tabular}{|c|c|c|c|c|c|c|}
\hline
 & $a_{calc.}$ & $(c/a)_{calc.}$ &$a_{exp.}$ [\onlinecite{bashkin}]  & $(c/a)_{exp.}$ [\onlinecite{bashkin}] &$a_{exp.}$ [\onlinecite{aksenenkov}] & $(c/a)_{exp.}$ [\onlinecite{aksenenkov}]\\
\hline &&&&&& \\
$\beta $      & 6.457   & 1     &   -   &   1     &   -   & -    \\
$\alpha$      & 5.860   & 1.583 & 5.866 &   1.583 &   -   & -    \\
$\omega$      & 9.122   & 0.617 & 9.152 &   0.617 & 9.131 & 0.616\\
\hline
\end{tabular}
\end{center}
\label{table:TiZ_lattice_parameter}
\end{table}

It is seen from the table that the lattice parameters $a$ and $c/a$ obtained in our calculation agree well with the experimental data. The greatest discrepancy is observed for the lattice constant of the $\omega$ phase. It should be noted that in Ref. \cite{bashkin} the lattice parameters were calculated for a metastable $\omega$ structure at atmospheric pressure, and in Ref. \cite{aksenenkov} for pressure-strained samples.
\begin{figure}[!tbh]
\resizebox{0.9\columnwidth}{!}{\includegraphics*{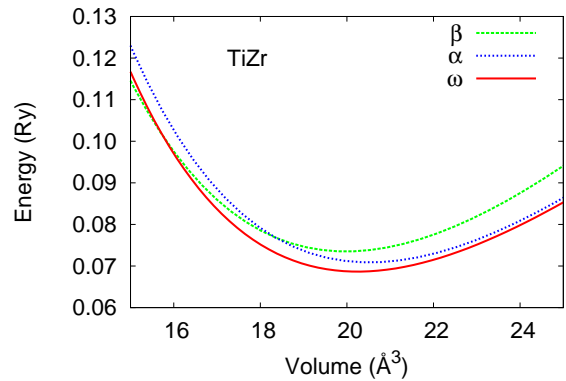}}
\caption{ Volume dependence of the total energy of equiatomic TiZr alloy for the $\beta$, $\alpha$ and $\omega$ phases}
\label{fig:TiZr_total_energy}
\end{figure}

The total energy of each structure was calculated for 7 values of the cell volume $V$. The data obtained were then interpolated using the technique proposed by Moruzzi \cite{Moruzzi}. Such an interpolation scheme, together with the Debye-Gr\"{u}neisen model, makes it possible to include implicitly anharmonic effects. The curves obtained for the volume dependence of the electron subsystem total energy are shown in Fig.~ \ref{fig:TiZr_total_energy}. The energy zero in figure corresponds to -8906.0 Ry.

As seen from the figure, the energy minimum in the ground state falls on the $\omega$ phase. And only at the relative volume change $V/V_0=0.75$ the $\beta$ phase becomes energetically preferable. A similar situation was observed in pure Ti and Zr as well. However the difference in energy between the $\alpha$ and $\omega$ phases in TiZr amounts to $\Delta E_{\alpha - \omega} = 6$~mRy, while in pure Ti and Zr it is $0.8$~mRy and $1$ mRy, respectively. Hence it follows that in the equiatomic alloy TiZr the stability region of the $\omega$ phase should be much larger  in temperature than in pure metals Ti,Zr. The equilibrium values of the volume are $V_{eq}= 20.47$ \AA$^3$, $20.52$ \AA$^3$, and $19.97$ \AA$^3$ for the $\alpha$, $\omega$ and $\beta$ phases, respectively.

\begin{figure}[!tbh]
\begin{center}
 \resizebox{1.0\columnwidth}{!}{\includegraphics*{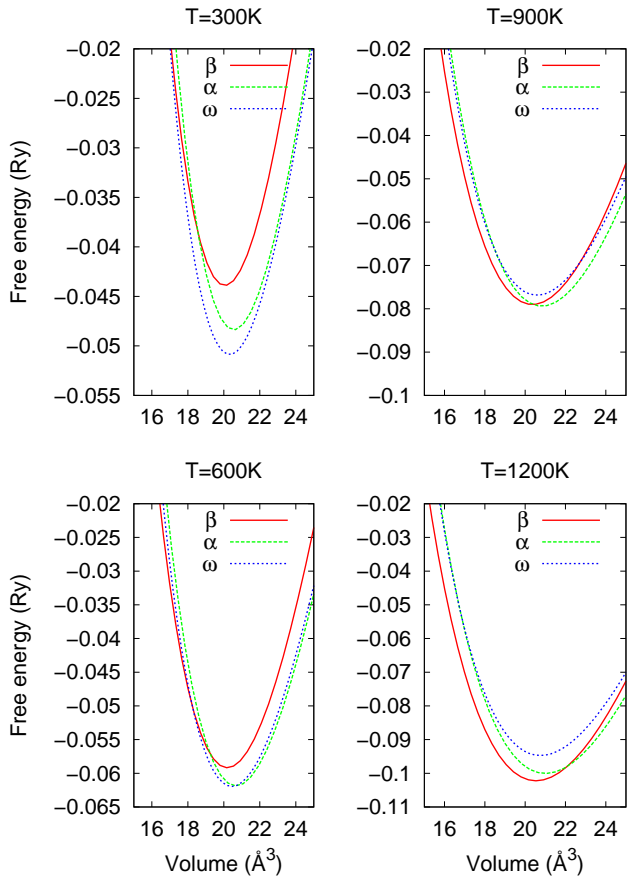}}
 \caption{Free energy of the $\alpha$, $\omega$ and $\beta$ phases of TiZr at different temperatures}
\label{fig:tizr_free}
\end{center}
\end{figure}
Figure \ref{fig:tizr_free} presents the volume dependence of the free energy at different temperatures. The free energy of the TiZr alloy was calculated in the Debye-Gr\"{u}neisen model with allowance made for the contributions from the electron entropy. The technique of calculating this latter has been described in detail in Ref. \cite{O.Eriksson-92}. As seen from the figure, the relationship between the energies of different structures changes with temperature. So, at $300$~K and zero pressure the energy minimum falls, as in the ground state, on the $\omega$ phase which remains energetically preferable up to $T=600$~K. In the temperature range $600$~K<T<$900$~K  it is the $\alpha$ phase of TiZr alloy which becomes stable, while above $900$~K it is the $\beta$ phase.

\section {Phase diagram}

\begin{figure}[!tbh]
 \resizebox{0.8\columnwidth}{!}{\includegraphics*{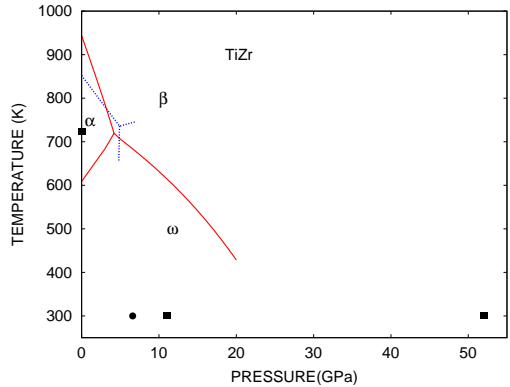}}
\caption{The P-T phase diagram of TiZr. The solid line shows the calculation results. The dotted lines are constructed from the experimental data [\onlinecite{bashkin}]. The experimental values for the  $\alpha - \omega$ transition at room temperature are taken from Refs.:  {\Large $\bullet$} - [\onlinecite{aksenenkov}], $\blacksquare$ - [\onlinecite{bashkin2}].}
\label{fig:tizr_Phase}
\end{figure}

The phase diagram of TiZr based on the analysis of Gibbs potentials for different structures is presented in Fig.\ref{fig:tizr_Phase}. The results of calculation are shown by the solid line. The dotted line denotes the experimental equilibrium boundaries for the $\alpha$, $\beta$ and $\omega$ phases of TiZr obtained in Ref. \cite{bashkin}. The experimental values of the  $\alpha - \omega$ transition at room temperature are taken from Refs. \cite{bashkin2,aksenenkov}.

On the whole, a good agreement of the calculated triple point ($P_{theor}=4.2$ GPa, $T_{theor}=720$ K) with the experimental values $P_{exp}=4.9 \pm 0.3$ GPa, $T_{exp}=733 \pm 30$ K \cite{bashkin} is observed. At zero pressure the calculated temperature of the $\beta - \alpha$ transition is  $T_{\beta - \alpha}^{theor}=943$  K. This value is higher than the experimental one, $T_{\beta - \alpha}^{exp}=852$ K, defined in Ref. \cite{bashkin} as the average of the temperatures of  the transition onset on heating and cooling. It should be noted that a large hysteresis is observed upon the $\alpha - \beta$ transformation in TiZr. At atmospheric pressure the maxima of thermal peaks in the DTA curves fall on $T \sim 912$ K on heating and $T \sim 810$ K on cooling, the typical peak width being $\Delta T \sim 40$ K. With this in mind, one can consider the results of calculation of the $\alpha - \beta$ equilibrium boundary in the Debye-Gr\"{u}neisen model as quite satisfactory.

The greatest discrepancy between the theoretical calculation and the experimental evidence available is observed for the $\alpha - \omega$ transition. In Ref. \cite{aksenenkov} the pressure at which this transition occurs at room temperature was estimated to be $P_{\alpha - \omega}^{exp}=6.6$ GPa.  Note thet equilibrium point of the $\alpha$ and $\omega$ phases was determined under shear-strain conditions at pressures up to $9$ GPa. The shearing strain is known to lower the pressure at which the phase transition begins. Presumably for this reason the authors of Ref.\cite{aksenenkov} failed to obtain the $\alpha - \omega$ transition at room temperature under quasi-hydrostatic conditions. In Ref. \cite{bashkin2} it was shown by X-ray diffraction method that the $\alpha$ phase of TiZr remains the sole stable phase under quasi-hydrostatic pressure up to $12.2$ GPa. Only from $5.5$ GPa on, becomes dominating the $\omega$ phase which remains stable up to $56.9$  GPa. At pressures above $56.9$ GPa there forms a high-pressure phase with a bcc lattice. 

As seen from Fig.\ref{fig:tizr_Phase}, in our calculation at atmospheric pressure and low temperatures the $\omega$ phase is stable, there occurs no $\alpha - \omega$ transition at room temperature. Note that in our calculations of pure Ti \cite{Ost-97} and Zr \cite{Ost-98}, in complete agreement with the experimental data, the $\alpha$ phase is stable at atmospheric pressure and room temperature, and the $\omega$ phase is stable only under pressure. That the $\omega$ phase in TiZr at normal conditions is energetically preferable, immediately follows from a comparison of the calculated free energies (see Fig.\ref{fig:tizr_free}).  Recall that the difference in energy between the $\alpha$ and $\omega$ structures in the equiatomic TiZr alloy is almost five times greater than in pure titanium and zirconium.

The discrepancy between experiment and theory may be due to the fact that the calculation was  performed for ideal crystalline structures (see Fig.\ref{Fig: TiZr_Cristal_structure}),whereas the experimental samples were imperfect crystals with lattice defects. In particular, it was shown \cite{aksenenkov} that in a TiZr alloy shear-strained under pressure the $\omega$ phase is represented by aggregations of oblong particles with characteristic size of $3-5$ nm, and $15-30$ nm long. If $\omega-$phase particles are situated in a coarse grain of $\alpha$ phase, they are mainly located at its boundaries. It was also noted \cite{bashkin} that various imperfect structures in samples pre-treated in different ways have a noticeable effect on the course of structural transformations. Evidently, we could not model a real structure in first-principles calculations.

The correctness of our results may be supported by the following experimental evidence \cite{bashkin}: firstly, the metastable $\omega$ phase was obtained at atmospheric pressure as a result of cooling of the $\beta$ phase under a pressure of 6 GPa with subsequent unloading at room temperature.  Secondly, between $2.2$ and $4.8$ GPa on cooling of the $\beta$ phase there forms a two-phase mixture of a stable $\alpha$ and a metastable $\omega$ phase. And lastly, it was found that at atmospheric pressure the $\omega$ phase in the TiZr alloy, when heated above $698$ K, transforms into an $\alpha$ phase \cite{bashkin}. In Ref. \cite{aksenenkov} the temperature of this transformation was defined as $T=623$ K for $P=0.0001$ GPa. This value differs by only $13$ K from the temperature $T_{P=0}^{\omega \to \alpha}=610$ K  we have calculated for the $\omega \to \alpha$ transition.
\begin{center}
\begin{figure}[!tbh]
 \resizebox{1.0\columnwidth}{!}{\includegraphics*{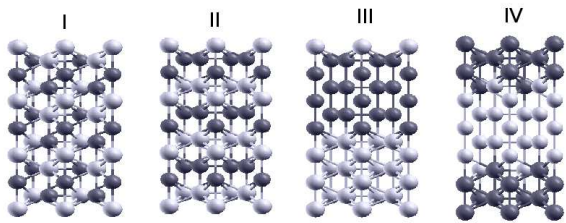}}
\caption{The structure types used in modeling the decomposition in $\omega-$TiZr. The Ti atoms are dark, the Zr atoms light.}
\label{Fig:rasl-reshetka}
\end{figure}
\end{center}

The above evidence suggests, in our opinion, that in the TiZr alloy with ideal crystal lattice the phase diagram should look as it is depicted in Fig.\ref{fig:tizr_Phase}. Of course, it must be taken into account that the Debye-Gr\"{u}neisen model, used for calculating the thermodynamic potentials, is a rather rough approximation and, obviously, cannot ensure good accuracy, especially at high temperatures when the anharmonic effects become of considerable importance. It should be noted that at room temperature the pressure calculated for the $\omega \to \beta$ transition is nearly half as large as the experimental value. Because of low temperatures, we do not believe this discrepancy to be connected with the choice of the Debye model for describing the thermodynamic properties. It is rather due to the deviation of real alloys from the ideal periodic structures used in calculating the total energy in the ground state.

\section {Calculation of the tendency toward decomposition}
To estimate the tendency toward decomposition in $\omega-$TiZr in the ground state, the total energy was calculated for four structures (see Fig.\ref{Fig:rasl-reshetka}). Structure I was represented by layers of pure titanium and pure zirconium alternating along the $z$ axis and separated by intermediate mixed Ti-Zr layers. In the two-layer structure II two layers of pure titanium alternated along the $z$ axis with two layers of pure zirconium with no intermediate layer. For structure III were chosen six Ti layers alternating with six Zr layers without intermediate layer. And lastly, in variant IV five Ti layers were separated by an intermediate layer from five Zr layers.

The free energy was calculated by the scalar relativistic full-potential linearized augmented-plane-wave (FPLAPW) method, using the WIEN2K package\cite{Wien2k}. In the first two variants the number of atoms per unit cell was six (3 Ti atoms and 3 Zr atoms). In variants III and IV the number of atoms in the cell amounted to 18 (9 atoms of each species). For variants I and II structural optimization of the $c/a$ ratio was performed, and the equilibrium atom positions were defined with the procedure of minimizing the forces acting on atoms.
\begin{center}
\begin{figure}[!tbh]
 \resizebox{0.8\columnwidth}{!}{\includegraphics*{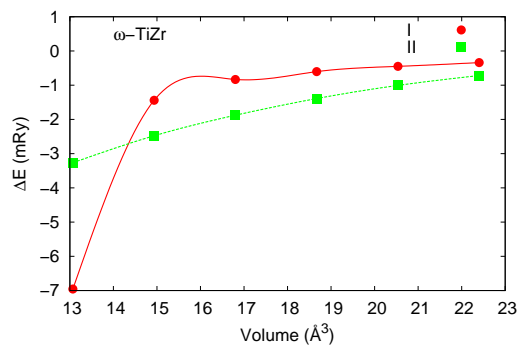}}
\caption{Energy change upon lattice relaxation with respect to the equilibrium atom positions in ideal $\omega$ phase for variants I and II}
\label{Fig:relax-energy}
\end{figure}
\end{center}

\begin{center}
\begin{figure}[!tbh]
 \resizebox{0.8\columnwidth}{!}{\includegraphics*{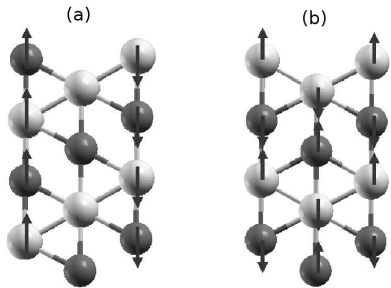}}
\caption{The direction of atomic displacements upon lattice relaxation for calculation variants I (a) and II (b)}
\label{Fig:shift-atom}
\end{figure}
\end{center}

\begin{center}
\begin{figure}[!tbh]
 \resizebox{0.8\columnwidth}{!}{\includegraphics*{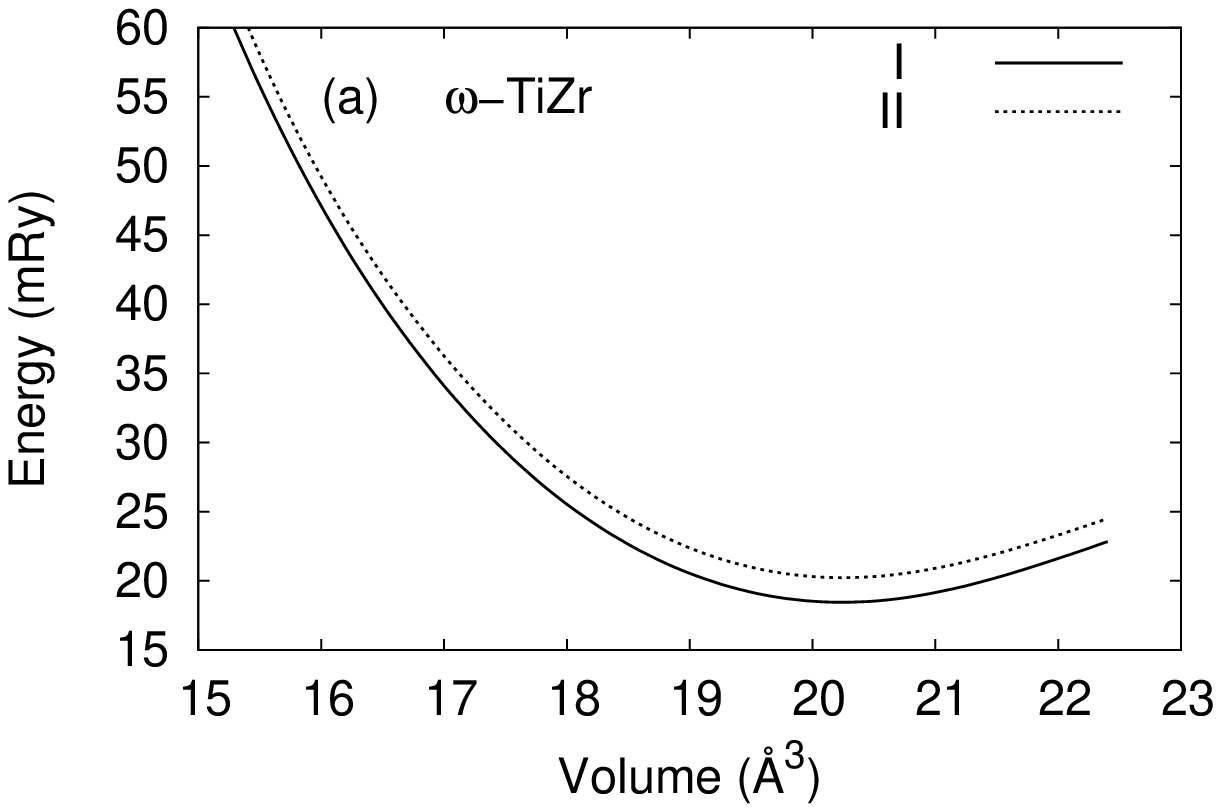}}
 \resizebox{0.8\columnwidth}{!}{\includegraphics*{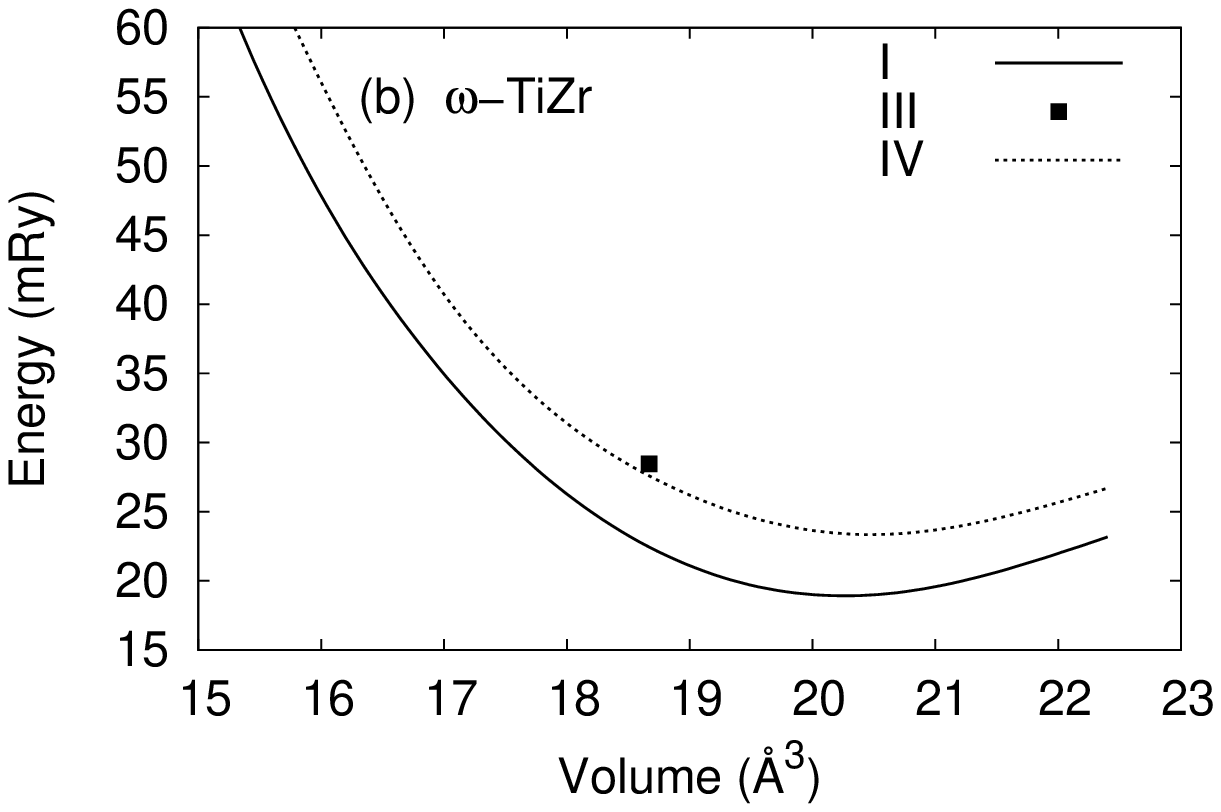}}
\caption{Total energy of $\omega-$ TiZr for different types of decomposition ( I - IV) designated
in accordance with Fig.\ref{Fig:rasl-reshetka}.}
\label{Fig:Total-energy-I-II-relax}
\end{figure}
\end{center}

In Fig.\ref{Fig:relax-energy} the volume dependence of the lattice relaxation energy change $\Delta E=E_{relax}-E_0$ is displayed for calculation variants I and II. Here $E_0$ is the total energy of the system with atomic arrangement corresponding to the ideal $\omega$ lattice; $E_{relax}$ is the system energy after minimization of the forces acting on atoms for a given volume. As seen from the figure, with decreasing volume the atoms become displaced from the positions corresponding to the ideal $\omega$ lattice, the displacement magnitude depending on the volume and the structure type. For the structure of type I, corresponding to the most uniform distribution of Ti and Zr atoms at $V<15$~\AA$^3$, there occurs a sharp decrease of $\Delta E$ due to significant atomic rearrangement. For the two-layer system (II) such a rearrangement is not observed in the considered  interval of volume change. In Fig.\ref{Fig:shift-atom}, for $V=13$~\AA$^3$ are depicted the (110) planes, the arrows indicating the direction of atomic displacements on relaxation for the lattices of type I (a) and II (b) (the $z$ axis is pointing upwards). 
It is seen that in both cases the atomic displacements are directed only along the $z$ axis. In the two-layer system (II) the atoms of titanium and zirconium are displaced in opposite directions, whereas in system I  the atomic chain displacement occurs without strain. The displacements shown in Fig.\ref{Fig:shift-atom}(a) correspond to those characteristic of the $\omega \to \beta$ transition. The volume value $V \approx 15$~\AA$^3$ at which begins a sharp decrease in $\Delta E$, agrees well with the results of the total energy calculation for the $\omega$ and $\beta$ phases of TiZr plotted in Fig.\ref{fig:TiZr_total_energy}. Based on the data presented, we can draw an important conclusion that the pressure value at which occurs the $\omega \to \beta$ transition depends substantially on the ordering type in the equiatomic TiZr alloy. This also indirectly confirms our statement that the disagreement with the experiment concerning the $\omega - \beta$ equilibrium boundary position on the phase diagram calculated in the Debye model (Fig.\ref{fig:tizr_Phase}) is caused by the presence of inhomogeneities in actual TiZr alloys used in experiments in Refs.\cite{bashkin,bashkin2,aksenenkov}.
The volume dependence of the total energy for the relaxed structures of type I and II is plotted in Fig.\ref{Fig:Total-energy-I-II-relax}(a). As may be seen, in the whole range of volume change the energy is minimum for the structure of type I corresponding to the most uniform distribution of Ti and Zr atoms. Recall that in variant I the lattice is represented by pure monolayers of titanium and zirconium separated by a mixed Ti-Zr layer, and in variant II by a system of two Ti layers alternating with two Zr layers  (with no intermediate layer). Thus, as decomposition grows, the system energy increases. This tendency persists on further decomposition, which may be seen in Fig.\ref{Fig:Total-energy-I-II-relax}(b), where the total energy is plotted versus volume for the structures of type II, III and IV. In the structure of type IV five Ti layers are separated from five Zr layers by a mixed intermediate layer, while in structure III there are six layers of each metal without intermediate layer. It is seen from the figure that as the thickness of pure metal layers increases (from one to five layers), the system energy significantly rises. As could be expected, the presence of an intermediate layer reduces the total energy of the system. This may be seen from a comparison of the energy values at $V \approx 18.5$~\AA$^3$ for the calculation variants III and IV (structure III is presented in the figure by a single point).
To summarize, the calculations performed show that in the ground state the $\omega$ phase of TiZr exhibits a tendency toward ordering and not toward decomposition, as was suggested in Ref.\cite{Bashkin-2008}. The analysis of the total energy curves shows that the allowance for temperature effects in the Debye-Gr\"{u}neisen model will not change the energy relation between different calculation variants, and cannot explain the experimentally observed formation  of two $\omega$ structures. Besides, we have not found any peculiarities connected with the s-d electron transition in the total-energy curves. Thus, also the suggestion advanced in Ref. \cite{Dmitriev} that there exists an isostructural transition $\omega \to \omega_1$ due to the pressure-induced changes in the electron structure, is not confirmed by the calculation. In our opinion, the high-temperature decomposition in the $\omega$ phase of equiatomic TiZr alloy is connected not with the change in electron structure under pressure, but with peculiarities of the lattice dynamics, in particular, with the presence of strongly anharmonic vibrational modes which are of crucial importance in stabilization of the $\omega$ lattice of pure titanium and zirconium\cite{Trubitsin}.

\begin{acknowledgments}
The authors acknowledge  the partial support from the  RFBR Grants No. 07-02-00973 and No. 07-02-96018.
\end{acknowledgments}

\end{document}